\documentclass[12pt]{article}
\usepackage[dvips]{graphicx}
\begin{document}
{\hfill{FTUV-00-0412}}
\vspace{2cm}

\begin{center}
{\Large{\bf RENORMALIZATION OF THE $f_0(980)$ and $a_0(980)$ SCALAR RESONANCES 
 IN A NUCLEAR MEDIUM \\
}}
\end{center}

\vspace{1.6cm}

\begin{center}
{\large{ E. Oset and M.J. Vicente Vacas}}
\end{center}

\begin{center}

{\small{ \it Departamento de F\'{\i}sica Te\'orica and IFIC, \\
Centro Mixto Universidad de Valencia-CSIC, \\
Institutos de Investigaci\'on de Paterna, Apdo. correos 2085,\\
46071, Valencia, Spain}}

\end{center}

\vspace{1cm}

\begin{abstract}
{\small{The meson meson interaction in the scalar sector in the presence of a
nuclear medium is studied with particular attention to the change of the 
properties of the $f_0(980)$ and $a_0(980)$ resonances. By using a chiral
unitary approach which generates the $f_0$ and $a_0$ resonances and reproduces
their free properties, we find that the position of the resonances in the
medium is barely changed but their widths are considerably broadened. New many
body corrections generating from higher orders in the chiral Lagrangian plus
the contribution from N*h excitations, not considered before in connection
with  the $\pi \pi$ interaction in the nuclear medium, are also investigated.}}
\end{abstract}

\vspace{2cm}


\newpage

\section{ Introduction}

 The scalar resonances produced in the scattering of pseudoscalar mesons have
been a permanent source of discussion\cite{montanet,PDG}.They are advocated  as
ordinary $q \bar{q}$ states \cite{tornqvist}, $q^2\bar{q}^2$ states 
\cite{jaffe}, $K\bar{K}$ molecules \cite{weinis, npa}, glueballs \cite{5deMP93}
and/or hybrids \cite{6deMP93}. The properties of these resonances, like their 
mass, width, partial decay widths and their influence in different reactions 
are closely tied to their nature, although different interpretations are
sometimes possible.
 
 The properties of the resonances are modified in the presence of a nuclear 
medium. They get a selfenergy which changes the position of the peak, the decay
width and in addition introduces new decay channels. The predictions  for the 
modification of these  properties are also closely related  to the  hypothesis
made on the nature of the resonance. On the other hand, the knowledge of the
renormalization of the resonance properties is an essential tool in order to
interpret correctly experiments  where pairs of pseudoscalar mesons are
produced in nuclei close to the energy  regions where the resonances appear. 
This would be the case for the production of two pions which can lead to the
$\rho$ resonance in L=1 and isospin I=1, or  the $\sigma$ and the $f_0(980)$ in
L=0 and I=0, or the production of $\pi\eta$  which can lead to the  $a_0(980)$
resonance in L=0 and I=1.
  
 In particular, the renormalization of the $\rho$ meson properties in nuclei 
has witnessed a spectacular effort both theoretical and experimental after the 
suggestion that there should be a universal scaling of the masses \cite{BR}. 
Although not supported experimentally\footnote[1] 
{The pion selfenergy is repulsive instead of attractive, as suggested for the 
$\rho$ in \cite{BR} on the basis of the value for the nucleon effective mass 
in the nucleus, which actually has a different meaning than the one attributed 
to the $\rho$ effective mass in that work.},
and also challenged theoretically by more recent calculations, the hypothesis
certainly had a stimulating effect which has lead to a series of thorough and 
detailed studies which have set the issue on firmer grounds 
\cite{friman, schuck, weise,wambach,mosel}. Most works would conclude that the 
$\rho$ mass does not appreciably change in the medium, but the width is 
substantially increased.  This latter change alone might be sufficient to 
explain the spectra of dileptons produced in heavy ion collisions 
\cite{cakorarho}, although further theoretical and experimental
work is being done.

  Another front where there has been much progress is the renormalization of 
the $\pi\pi$ scattering amplitude in a nuclear medium and its possible 
relationship to the enhancement of the $\pi\pi$ invariant mass distribution 
close to threshold, seen in the experiments of pion induced two pion production
in nuclei \cite{nevio1}. It was suggested in \cite{chanfray} that the 
$\pi \pi$ interaction could develop a singularity just below the two pion 
threshold which would correspond to a kind of Cooper pair. Since this shows up 
in the scalar and isoscalar channel it was also interpreted as a drop of the 
sigma mass in the nuclear medium. More refined calculations which have payed 
special attention to the chiral constraints of the $\pi\pi$ amplitude 
appreciably weaken the renormalization of the amplitude, although there
is still an appreciable enhancement of the imaginary part close to threshold
\cite{rapp,aouissat, chiang}. Although there were initial hopes that this alone
could explain the experimental features \cite{rapp2} of the $(\pi,\pi\pi)$ 
reaction in nuclei, more detailed calculations have shown that some additional 
mechanisms may be required \cite{manolo}. 

   The former examples show two fronts where the medium properties of mesons
are thoroughly investigated. In the present paper we want to pay attention to
the modification of the meson scalar resonances  $f_0(980)$ and $a_0(980)$
inside the nuclear medium. The $f_0(980)$ resonance has the same quantum numbers
as the $\sigma$, only its mass is much larger, and conversely the width is much
smaller, of the order of 40-100 MeV (versus $m_\sigma=450 MeV$ and a width 
of around 450 MeV according to the PDG book \cite{PDG}). The $a_0(980)$ in the 
I=1 channel has about the same mass as the $f_0(980)$ and a width of the order
of 50-100 MeV. 

In spite of its relevance as a source of information on the nature of these 
resonances and the interest that it should have in the analysis of the two 
meson production in nuclei, the surprising fact is that there are neither 
theoretical nor experimental studies on this issue, which contrast with the 
large efforts devoted to the study of the $\rho$ meson properties in an energy 
region very close to the one where these resonances appear. The reasons for 
this might be simply technical. On the one hand the confusion about the nature 
of these states was a deterrent. On the other hand, the study of the properties
of these resonances in nuclei, which couple both of them largely to the 
$K \bar{K}$ system, had their own problems since the question of the $K^-$ 
selfenergy in a nuclear medium was itself unclear \cite{gal,koch2,waas,lutz}. 
Fortunately things have changed in both fronts recently, to the point that 
one can count on reasonable models with which to tackle the problem. 

 One of the areas which has witnessed an important progress in recent years is
the meson meson interaction by means of Chiral Perturbation Theory 
($\chi PT$), which is supposed to be the effective theory of QCD at low energies
\cite{weinberg}. The theory has proved rich in applications to strong, weak
and electromagnetic processes in which pairs of mesons appear at small energies 
\cite{weinberg,pich,meissner,eckerep}. Yet, implicit to the perturbative 
nature of the theory is the fact that it does not generate poles in the 
scattering amplitudes and hence is unsuited to study the energy regions where 
meson resonances appear. In this respect, there have been recent advances which 
have shown the usefulness of $\chi PT$ as a means to constrain non perturbative
unitary methods. For instance, by means of the inverse amplitude method (IAM) 
\cite{IAM} and the chiral Lagrangians, one can obtain the $\sigma$ and $\rho$ 
mesons in the $\pi\pi$  scattering and the $K^*$ resonance in $K\pi$  
scattering. The generation of the $f_0(980)$ and $a_0(980)$ resonances, however,
required the extension of the IAM to coupled channels, which was done in 
\cite{prl,prd,npb} where these two mesons were also generated. In addition, 
all meson meson properties up to about 1.2 GeV were reproduced using simply 
the standard $O(p^2)$ and $O(p^4)$ chiral Lagrangians as input.

  Advances have also been made using the hypothesis of resonance
saturation of Ref. \cite{ecker}, which states that the information of the 
$O(p^4)$ chiral Lagrangian is tied to the exchange of resonances which survive
in the large $N_c$ limit. By allowing such genuine (preexisting to the
unitarization, or multiple scattering of the mesons) resonances and using the 
$O(p^2)$ chiral Lagrangian in addition, together with  a proper unitarization 
scheme based on the N/D method, it was shown in \cite{oond} that a good 
description of all the meson meson information up about 1.5 GeV could be 
accomplished. In this respect it is interesting to observe that the use of the 
lowest order $\chi PT$ Lagrangian, properly unitarized by means of the Bethe 
Salpeter equation, together with an appropriate cut off to regularize the loops,
is able to reproduce all the information of the meson meson scattering in the 
scalar sector up to about 1.2 GeV and gives the same results as the more 
general methods reported above. This is possible due to the large weight of the
lowest order Lagrangian in the scalar sector, in contrast to what happens in the
vector sector where higher orders play an essential role.

   The meson baryon interaction, from the point of view of $\chi PT$, has also
witnessed much progress \cite{pich, meissner, eckerep,nadia}. On the other hand
the meson baryon interaction has  been the ground for application of the
chiral unitary techniques, using the Lippmann Schwinger summation
\cite{norbert1,norbert2} or the Bethe Salpeter integral equation \cite{ramos}
and more recently the IAM in \cite{ramonet} or the N/D method
\cite{ulfoller}.

  The work of \cite{norbert1,norbert2} initiated a fruitful line which has
allowed to link the $K^- N$ interaction with its coupled channels at low
energies, plus the properties of the $\Lambda(1405)$ resonance, with the
information of the chiral Lagrangians. The work of \cite{ramos} added more
channels in the coupled channels set and proved that it was possible to get a
good reproduction of the low energy data in terms of the lowest order chiral
Lagrangian and a suitable cut off for the loops. 

    This progress in the elementary reaction also
stimulated work directed to obtain the selfenergy of the kaons in nuclei. It was
soon realized that the consideration of Pauli blocking in the intermediate
nucleon states lead to a shift towards higher energies of the $\Lambda(1405)$
resonance and, as a consequence, to an appreciable attraction on the kaon
\cite{koch2,waas}. However, the large kaon attraction which was found suggested
that the problem had to be solved selfconsistently. This was done in 
\cite{lutz}, where it was found that the selfconsistent treatment rendered the 
resonance to the same free position, although it appeared with a substantially 
larger width. This problem has been further pursued in \cite{angoset} where all 
elements of \cite{lutz} were considered but the renormalization of the pions 
in the intermediate states was also taken into account. With all these 
ingredients a $K^-$ nucleus optical potential was found which has proved 
consistent with the information of kaonic atoms \cite{hiren}.  

   Thus, the situation at present is that one has the information and
the means to generate the scalar mesons from the chiral Lagrangians and also the
interaction of the kaons with the nucleus, the two ingredients that blocked
progress in the topic of the scalar meson renormalization in nuclei, and this is
the problem that we shall tackle in this paper.

\section {The meson meson  interaction in the nuclear medium}

  Here, we briefly review the formalism for the meson meson interaction which was
developed in \cite{chiang}. We start by taking the states 
$\pi \pi$ and $K \bar{K}$ ,which we label 1 and
 2. In the I=0 channel they are given by
\begin{equation}
\begin{array}{l}
| K \bar{K} > = - \frac{1}{\sqrt{2}} 
 | K^+ (\vec{q}) K^- (- \vec{q}) + K^0 (\vec{q}) \bar{K}^0 (- \vec{q})>\\[2ex]
| \pi \pi >  = - \frac{1}{\sqrt{6}}
| \pi ^+ (\vec{q}) \pi^- (- \vec{q}) 
+ \pi^- (\vec{q}) \pi^+  (- \vec{q}) + \pi^0 (\vec{q})  \pi^0 (- \vec{q})>,
\end{array}
\end{equation}
and the  $K \bar{K}$, $\pi \eta$ which we label 1 and 2, in the isospin I=1
channel,
\begin{equation}
\hspace{-3cm}
\begin{array}{l}
| K \bar{K} > = - \frac{1}{\sqrt{2}}
 | K^+ (\vec{q}) K^- (- \vec{q}) - K^0 (\vec{q}) \bar{K}^0 
(- \vec{q})>\\[2ex]
| \pi \eta >  = |\pi ^0 (\vec{q}) \eta (- \vec{q})>,
\end{array}
\end{equation}
where $\vec{q}$ is the momentum of the particles in the CM of the pair. 
We follow the convention $| \pi^+ > = - |1,1 >$ and 
$|K^- > = -|\frac{1}{2}, -\frac{1}{2} > $ isospin states.
  In the I=0 case we neglect the $\eta \eta $ channel. Its contribution has 
been assessed in \cite{oond,norberteta} and it has only relevance
at energies beyond 1.2 GeV, hence, as done in \cite{npa}, we shall omit this
channel here too.
 
    The Bethe Salpeter (BS) equation is given by 
\begin{equation}
\label{eq:BS}
T=VGT.
\end{equation}   
Eq. \ref{eq:BS} is meant as a coupled channel equation with two channels in each
of the isospin states. It is an integral equation, meaning that the term $VGT$ 
involves one loop integral
where both $V$ and $T$ would appear off shell. However, it was shown in
\cite{npa,arriola} that, for the free case, these amplitudes could be factorized
on shell out of the integral and the off shell part was absorbed by a 
renormalization of the coupling constants. 
 Thus, the equation becomes purely algebraic and is quite easy to solve. In Eq.
\ref{eq:BS}, the formal product of
$VGT$ inside the loop integral becomes then the product of $V$,$G$ and $T$, 
with $V$ and $T$ the on shell amplitudes, and   
the function $G$ is given by  the diagonal matrix  
\begin{equation}
G_{ii} = i \int \frac{d^4 q}{(2 \pi)^4}
\frac{1}{q^2 - m_{1i}^2 + i \epsilon} \; \; 
\frac{1}{(P - q)^2 - m_{2i}^2 + i \epsilon}
\end{equation}
where $P$ is the total fourmomentum of the meson-meson system.

\begin{figure}[htb]
 \begin{center}
\includegraphics[height=2.cm,width=12.cm,angle=0] {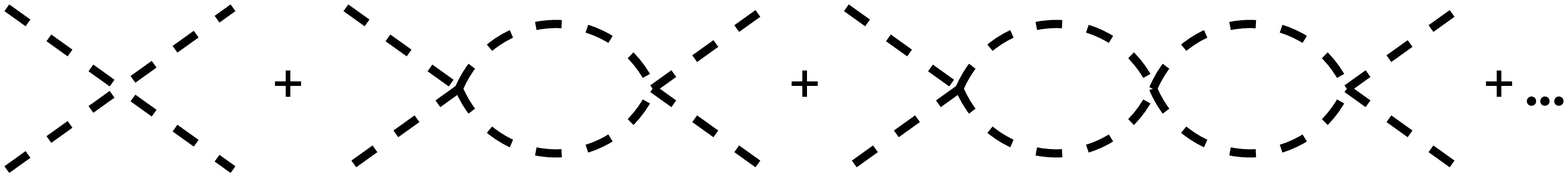}
 \caption{Diagrammatic representation of the Bethe-Salpeter equation.}
 \label{fig:BSF}
 \end{center}
\end{figure}

\begin{figure}[htb]
 \begin{center}
\includegraphics[height=2.2cm,width=12.2cm,angle=0] {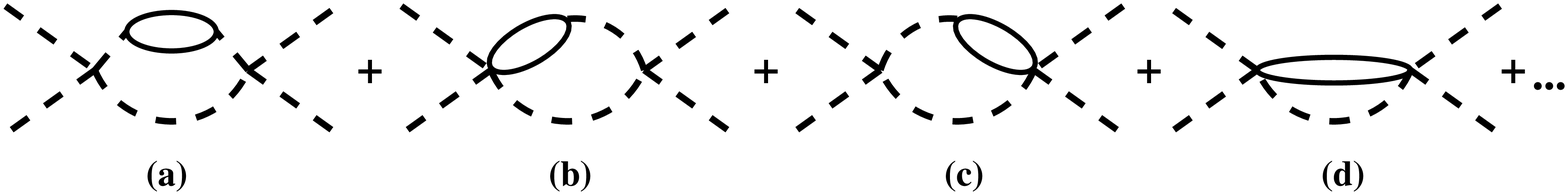}
\caption{Terms of the meson-meson scattering amplitude accounting for
$ph$ and $\Delta h$ excitation.}
 \label{fig:BSF2}
 \end{center}
\end{figure}

The BS equation sums the series of diagrams
depicted in Fig.~\ref{fig:BSF}.
In the nuclear medium, one has
to add the diagrams depicted in Fig. \ref{fig:BSF2}, which stem from the
interaction of the pions with the medium, through $ph$ and $\Delta h$ 
excitation. We neglect here the small s-wave pion-nucleon interaction.  
The interesting
finding of \cite{davesne} and \cite{chiang} was that the contact terms involving
the $ph$ ($\Delta h$) excitations of diagrams (b)(c)(d) canceled exactly the 
off shell contribution  from the meson meson vertices in the term of Fig.
\ref{fig:BSF2}(a). 
Hence technically one only had to evaluate the diagrams of the free type 
(Fig. \ref{fig:BSF}) and those of Fig. \ref{fig:BSF2}(a)
 (plus higher order iterations ) in order to evaluate the
pion pion scattering amplitude in the nuclear medium.  

In \cite{chiang}, only the pions were renormalized inside the medium since 
the work was only concerned with the meson-meson interaction at low energies 
where the kaons do not play much of a role. However, in order to address now 
the question  of the $f_0(980)$ and $a_0(980)$ resonances  we shall have 
to take into account the kaon  selfenergy in the medium. Since we are close to 
the two kaon threshold, the s-wave part of the kaon selfenergy  is the relevant 
ingredient.  For completeness, we shall also consider the
p-wave selfenergy due to $\Sigma h$ or $\Lambda h$ excitation, as in
\cite{angoset}, although it does not play an important role in the present
problem.

The loop involving pions is done as in \cite{chiang}. This provides the 
$\pi \pi$ matrix element of the $G$ function in the medium.
As for the $K \bar{K}$ intermediate  states we have an asymmetric situation.
In particular, the selfenergy of the $\bar{K}$ requires special care. 
We will use the results of the detailed model of Ref. \cite{angoset}. 
On the other hand, the $K$ selfenergy can be accounted for in a much 
simpler way since there are no resonances with strangeness $S=1$ and the 
$KN$ interaction is quite smooth. Altogether, $t\rho$ gives a very reasonable 
approximation to the $K$ selfenergy.  By taking results from \cite{waas} 
or \cite{ramos} we can write
\begin{equation}
\Pi=\frac{1}{2}(t_{Kp}+t_{Kn}) \rho \approx 0.13 m_{K}^2 \frac{\rho}{\rho_0}
\end{equation}
where $t$ is the elastic $K$-nucleon amplitude, $\rho$ is the nuclear
density and $\rho_0$ is the normal nuclear density.

  The new meson-meson amplitude in the medium is now given by means of the
modified BS equation 
\begin{equation}
\label{eq:BSN}
\tilde{T}=V\tilde{G}\tilde{T},
\end{equation}   
where the $K\bar{K}$ matrix element of $\tilde{G}$ is given by
\begin{equation}
\tilde{G}_{K\bar{K}} = i \int \frac{d^4 q}{(2 \pi)^4}
\frac{1}{q^2 - m_{K}^2-\Pi_{K}(q^0,q,\rho)        } \; \; 
\frac{1}{(P - q)^2 - m_{K}^2 -\Pi_{\bar{K}}(P^0-q^0,q,\rho)}
\end{equation}
The $\bar{K}$ propagator can be written in the Lehmann representation,
\begin{equation}
\frac{1}{(P - q)^2 - m_{K}^2 -\Pi_{\bar{K}}(P^0-q^0,q,\rho)}=
\int_{0}^{\infty}d \omega 2 \omega \frac{S_{\bar{K}}(\omega,q,\rho)}
{(P^0 - q^0)^2-\omega^2+i\epsilon}
\end{equation}
where $S_{\bar{K}}(\omega,q,\rho)$ is the  spectral function of the $\bar{K}$ 
in the medium, as described in \cite{angoset}. The $q^0$ and angular 
integrations can be done analytically and then we find  
\begin{equation}
\tilde{G}_{K\bar{K}} =   \frac{1}{2 \pi^2}\int_{0}^{\infty}d \omega
\int_{0}^{q_{max}}dq\, q^2  S_{\bar{K}}(\omega,q,\rho)
\frac{\omega+\tilde{\omega}(q)}{\tilde{\omega}(q)
(s-(\omega+\tilde{\omega}(q))^2) + i\epsilon}
\end{equation} 
where $\tilde\omega(q)$ is given by
\begin{equation}
\tilde\omega(q)=\sqrt{\vec{q}\,^2+m_{K}^2+\Pi_{K}}.
\end{equation} 
  For the $\pi \eta$ intermediate channel we proceed in the same way. The $\pi$
propagator is also written in  terms of the Lehmann representation, as done in
\cite{chiang} and the $\eta$ propagator explicitly in terms of the eta 
selfenergy, as we have done above for the $K$. We also take the $t\rho$ 
approximation for the $\eta$ selfenergy. However, the $t$ matrix is not so well 
known in this case. We take the following amplitudes from the fit of 
\cite{wycech} 
\begin{equation}
t^{-1}=-\frac{M}{4\pi\sqrt{s}}(\frac{1}{a}+r_0 q_\eta^2+s_0 q_\eta^4-i q_\eta),
\end{equation} 
where $M$ is the nucleon mass, $\sqrt{s}$ is the center of mass energy of the
$\eta$-nucleon system, and $q_\eta$ is the momentum of the $\eta$ meson
in the same system, assuming $\eta$ and nucleon to be on shell,
$a=0.75+ 0.27i\, fm $, $r_0= -1.50- 0.24 i\, fm$ and $s_0=-0.10-0.01 i\, fm^3 $.
We use these results from $\eta N$ threshold up to a value of $\sqrt{s}$ 200 MeV
above it. Outside the region of validity of this fit, we have assumed a 
constant selfenergy equal to the one of the closest extreme of the parametrized
region. The uncertainties associated to this assumption are small since taking
just a cero $\eta$ selfenergy outside that range of energies changes the results
by less than 5{\%}.

\section{Other medium corrections}
\subsection{Tadpole terms}
\begin{figure}[htb]
 \begin{center}
\includegraphics[height=2.5cm,width=5.cm,angle=0] {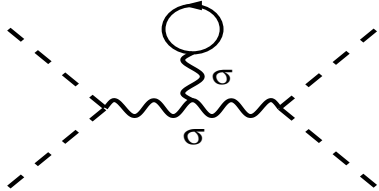}
 \caption{ Tadpole $\sigma$ selfenergy diagram.}
 \label{fig:SIG}
 \end{center}
\end{figure}

\begin{figure}[htb]
 \begin{center}
\includegraphics[height=2.0cm,width=12.5cm,angle=0] {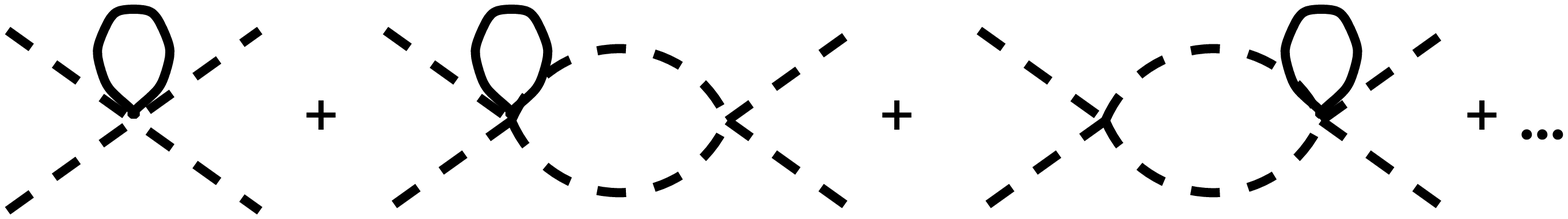}
 \caption{ Some tadpole diagrams contributing to the meson-meson scattering 
 amplitude }
 \label{fig:SIGPT}
 \end{center}
\end{figure}

In this section we consider some additional many body corrections. 
We begin by the contribution of the diagram depicted in Fig. \ref{fig:SIG}.
which has been considered in \cite{hatsuda}  using the linear sigma model,
with sigmas and pions as elementary fields. The Lagrangian gives rise to a
vertex with three sigmas as shown in the figure which produces a many body
correction, assuming a certain coupling of the sigma to the nucleons, which in
\cite{hatsuda}
is borrowed from the Bonn phenomenological boson exchange models of the $NN$
interaction \cite{karl}. On the other hand, the  $\chi PT$ Lagrangian involves
only pseudoscalar meson fields. The sigma can be generated dynamically  
through the rescattering of the pions. The closest analog to the diagram of 
Fig. \ref{fig:SIG}  is given by the series of Fig. \ref{fig:SIGPT} where the
sigma is generated to the left and right of the nucleon-hole loop by means of
the iteration of the chiral Lagrangian.
 In order to evaluate the contribution of these diagrams we use the
chiral Lagrangians involving the octet of baryons and the octet of 
pseudoscalar mesons\cite{pich,meissner,eckerep}. The terms needed come 
from the covariant derivative terms of the Lagrangian. After some trivial 
algebra, we find that for the scalar channel these terms are proportional to 
$\bar{p}\gamma^\mu p-\bar{n}\gamma^\mu n$ and therefore would vanish in 
symmetric nuclear matter after summation over protons and neutrons.
  
\begin{figure}[htb]
 \begin{center}
\includegraphics[height=3.0cm,width=5.5cm,angle=0] {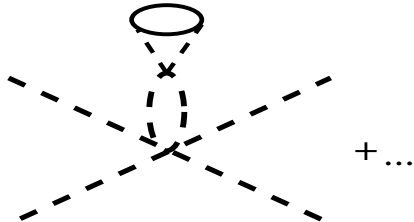}
 \caption{Higher order tadpole diagram. }
 \label{fig:SIGPT2}
 \end{center}
\end{figure}

 Considering higher order corrections, there is another possible way to 
generate an analogous structure as it is shown in Fig. \ref{fig:SIGPT2}
 
Here one also needs the coupling of the sigma, generated through pion-pion
interaction,  with the nucleons.  The scalar isoscalar exchange in the $NN$
interaction has been addressed in \cite{rolf,gestern} only at the perturbative
level. In \cite{toki} it has been revisited taking into account all the meson
meson rescattering which generates the sigma. In this latter work one finds an
attraction (not generated in the perturbative approach) at intermediate
distances, but weaker than  in the Bonn model because an appreciable repulsion
sets up already at distances like 0.7 fm, which grows fast at small distances.
However, the relevant magnitude would be the strength of the potential in
momentum space at zero momentum, which is what is met in Fig. \ref{fig:SIGPT2} 
and there the
strength of the potential of \cite{toki} is about one order of magnitude
smaller than for the Bonn potential.  Hence, from this source we
also get a negligible contribution to the modification of the $\pi \pi$
scattering in the nuclear medium. 

\subsection{Roper-hole excitation}
\begin{figure}[htb]
 \begin{center}
\includegraphics[height=3.0cm,width=5.5cm,angle=0] {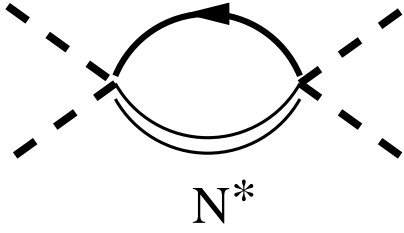}
 \caption{First order resonance-hole contribution to the meson meson scattering.}
 \label{fig:ROP1}
 \end{center}
\end{figure}

Finally, we will consider the excitation of resonances from the occupied 
nucleon states by the pair of mesons. This requires, in our case, a resonance 
which decays into a nucleon and two pions (two mesons in general) in
s-wave and with I=0. 
There is such a candidate at the energies where we are
concerned which is the $N^*(1440)$ Roper resonance.  The branching ratio for
this decay is small but it plays a very important role in the  
$\pi N \rightarrow \pi \pi N$ reaction 
\cite{tesina,Sossi:1993zw,dillig,ulf2pi,miranda}. It has also been
shown that this  mechanism is the dominant in the $N N \rightarrow N N  \pi
\pi $ reaction close to threshold \cite{luis,Alvarez-Ruso:1999xg}. 
In the present problem it could contribute via the mechanism depicted
in Fig. \ref{fig:ROP1}.

The evaluation of this mechanism is straightforward in the $SU(2)$ sector. In 
the first place we take the Lagrangian suggested in \cite{ulf2pi} which reads
\begin{equation}
{\cal L}_{N^* N \pi \pi} = c^*_1
\bar{\psi}_{N^*} \chi_+ \psi_N - \frac{c^*_2}{M^* \,^2 }(\partial_\mu
\partial_\nu \bar{\psi}_{N^*} ) u^\mu u^\nu \psi_N + h. c.
\end{equation}

\noindent
where

\begin{equation}
\chi_+ = m_{\pi}^2 (2 - \frac{\vec{\phi}\,^2}{f^2} + \ldots ) \, ; \, u_\mu = 
- \frac{1}{f} \vec{\tau} \partial_\mu \vec{\phi} + \ldots
\end{equation}

\noindent
and $\vec{\phi}$ is the pion field.
  The value of the $c^*_1$ and $c^*_2$ constants is not fully determined by
the partial decay width of the $N^*$, which only tells us that these 
values satisfy the equation of an ellipse. One needs to find extra constraints 
and in \cite{luis} it was found that the best set that lead to agreement with 
the $\pi^- p \rightarrow \pi^+ \pi^- p$ reaction was $c^*_1= -7.27 GeV^{-1}$ 
and $c^*_2$=0. This was corroborated in \cite{manolo} by looking at all the
different isospin channels. With the $c^*_1$ term alone we obtain the 
Lagrangian, after we expand up to two pion fields

\begin{equation}
{\cal L}_{N^* N \pi \pi} = -\frac{1}{f^2}c^*_1 m_{\pi}^2
          (\pi^0\pi^0+2 \pi^+\pi^-)(\bar{p}^*p+\bar{n}^*n) + h. c.
\end{equation}

There is no experimental information on the $N^*N K \bar{K}$ coupling. For
simplicity we will assume the simple generalization of the previous Lagrangian
\begin{equation}
{\cal L} = \frac{1}{2}c^*_1 \, Tr(\bar{B}B) Tr( \chi_+) \, ,
\end{equation}
The baryonic matrix $B$ and the mesonic matrix  $\chi_+$ can be found in
Ref. \cite{meissner}.
Expanding this Lagrangian up to terms with  two meson fields we obtain
and keeping only those terms relevant to our calculation we get 
\begin{equation}
\label{eq:lag1}
{\cal L} = -\frac{c^*_1}{f^2} \{ 
                         m_{\pi}^2 (\pi^0\pi^0+2 \pi^+\pi^-)
	               + m_{K}^2 (K^0\bar{K}^0+K^+K^-)
	                        \}
	  (\bar{p}^*p+\bar{n}^*n) + h. c.
\end{equation}
 We can see that the strength of the coupling to $K \bar{K}$ is $\frac{m_K^2}
{m_\pi^2}$ times that of the charged pions. Other simple generalizations of the
$SU(2)$ Lagrangian, i.e. $Tr(\bar{B}\chi_+B)$, also provide that scaling,
although with some different numeric factors and charge combinations.
With the choice of Eq. \ref{eq:lag1}, the contribution of the mechanism of 
Fig.\ref{fig:ROP1} to the meson meson  scalar isoscalar interaction is then 
given by
 
\begin{equation}
  V_{ij(N^*N)}' = (c^*_1)^2 
  \frac{4 m_i^2 m_j^2}{f^4} {\cal U}(p^0,\vec{p}=0,\rho)
\end{equation}
where $i,j$ stand for the physical meson states, 
$m_i, m_j$ are the masses of the initial and final mesons and
${\cal U}(p^0,\vec{p}=0,\rho)$ is the complex Lindhard function for the 
excitation of resonances , taken from \cite{salcedo}, which in the limit of 
$\vec{p}$=0,  which one has for a meson pair in their center of mass frame, 
has the simple expression 
\begin{equation}
 {\cal U}(p^0,\vec{p}=0,\rho)= \frac{\rho}{p^0-m_{N^*}+m_N+
                             i \frac {\Gamma_{N^*}}{2}}.
\end{equation}
This leads to an additional contribution to the $\pi \pi$ "potential" used 
in the Bethe Salpeter equation which is given by the matrix elements for I=0 
(for I=1 one gets zero contribution)
\begin{eqnarray} 
V'_{11} &=& 2 (c^*_1)^2 \frac{4 m_K^4}{f^4} {\cal U}(p^0,\vec{p}=0,\rho)\\
V'_{12} &=& \sqrt{3} (c^*_1)^2 \frac{4 m_\pi^2 m_K^4}{f^4} 
                        {\cal U}(p^0,\vec{p}=0,\rho)\\
V'_{22} &=& \frac{3}{2}(c^*_1)^2 \frac{4 m_\pi^4}{f^4} 
          {\cal U}(p^0,\vec{p}=0,\rho)
\end{eqnarray} 
Now one must be cautious to use the empirical value of  the $c^*_1$ parameter
when the potential of the former equations is put into the kernel of the BS 
equation. Indeed, at the level of terms linear in the density one will generate
the diagrams of the figure
\begin{figure}[htb]
 \begin{center}
\includegraphics[height=2.5cm,width=10.5cm,angle=0] {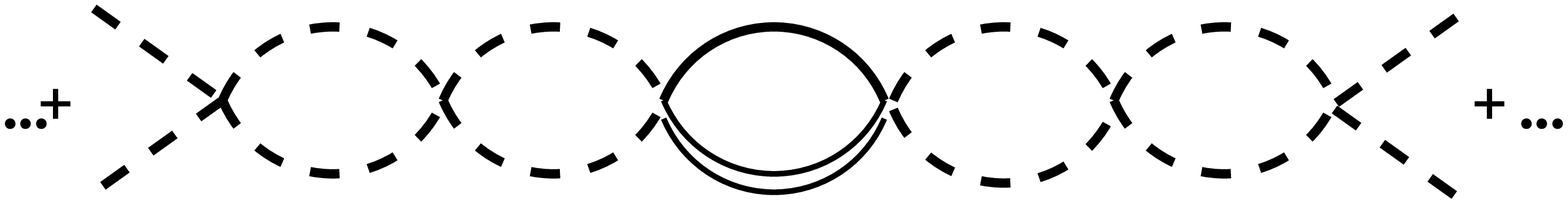}
 \caption{Higher order resonance-hole contribution to the meson meson scattering.}
 \label{fig:ROP2}
 \end{center}
\end{figure}

 This means that we are generating the iteration of the free mesons to the right
 and the left of the $N^*h$ excitation. As a consequence the empirical coupling
 should be related to a bare one  by
\begin{equation} 
c^*_1=c^*_{1B} (1+G_{\pi\pi}T_{\pi\pi,\pi\pi}^{I=0}+\sqrt{\frac{4}{3}}
\frac{m_K^2}{m_\pi^2}G_{KK} T_{K\bar{K},\pi\pi}^{I=0})
\end{equation}
  Thus the calculations must be done using the bare coupling, since the free
pion interaction renormalizes it to the effective values demanded by the $N^*$
decay into two pions in s-wave.

\section{Results and discussion}

\begin{figure}[htb]
\begin{center}
\includegraphics[height=12.cm,width=12.cm,angle=0] {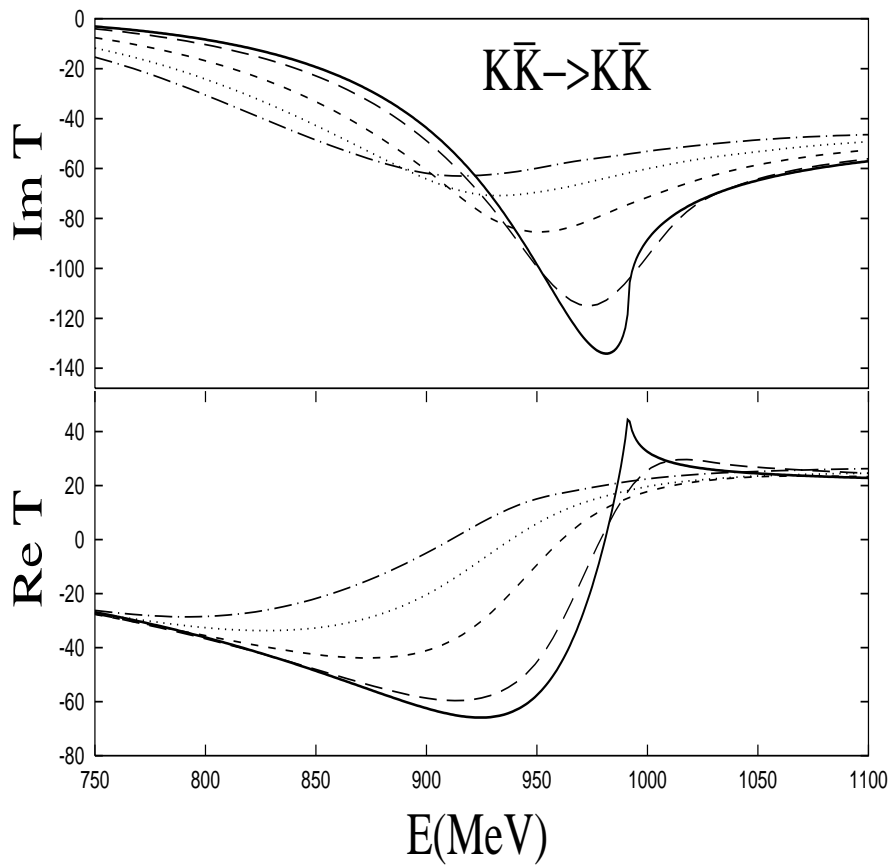}
\caption{Real and imaginary part of the kaon-antikaon scattering amplitude 
in the I=1 channel as a function of the center of mass energy for different 
nuclear densities. Solid line, free amplitude; long dashed line, 
$\rho=\rho_0/8$; short dashed line, $\rho=\rho_0/2$; dotted line, 
$\rho=\rho_0$; dashed dotted line, $\rho=1.5\rho_0$.}
\label{fig:R1}
\end{center}
\end{figure}

\begin{figure}[htb]
 \begin{center}
\includegraphics[height=12.cm,width=12.cm,angle=0] {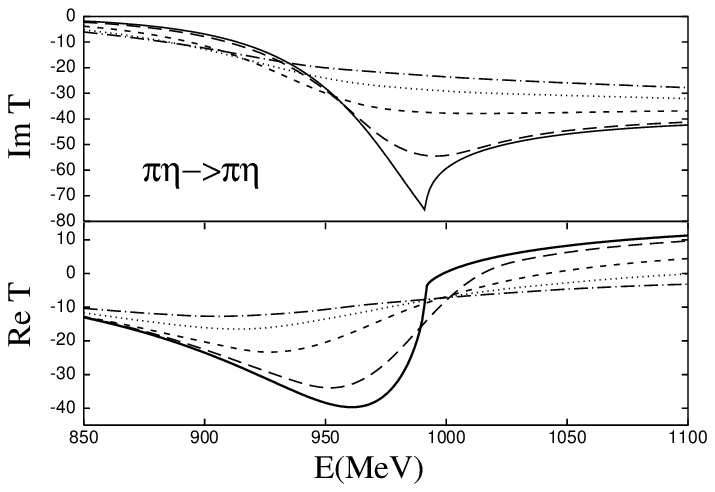}
 \caption{Same as Fig. 8 for $\pi\eta$ scattering.}
 \label{fig:R2}
 \end{center}
\end{figure}

\begin{figure}[htb]
 \begin{center}
\includegraphics[height=12.cm,width=12.cm,angle=0] {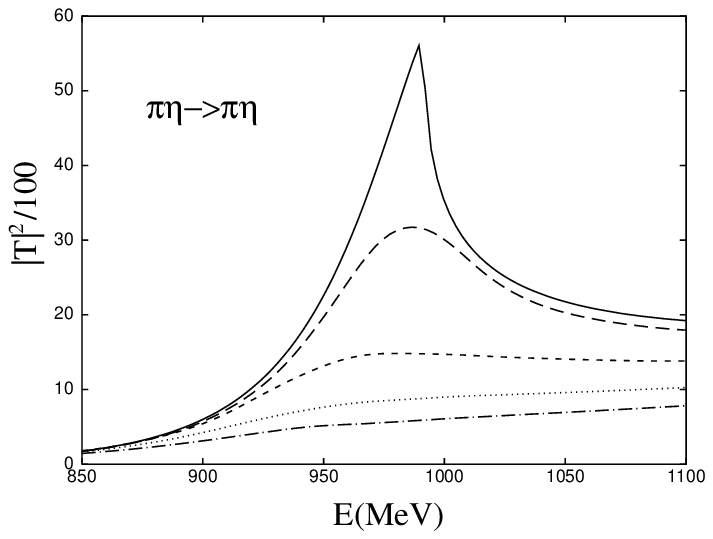}
 \caption{Modulus squared of the $\pi\eta\to\pi\eta$ amplitude for different
 densities. Meaning of the lines as in Fig. 8.}
 \label{fig:R3}
 \end{center}
\end{figure}

 In the first place let us discuss the results in the I=1 channel. In figs.
\ref{fig:R1},\ref{fig:R2} we show the real and imaginary parts of the 
$K \bar{K} \to K \bar{K}$ and
$\pi \eta \to \pi \eta $ amplitudes for different values of the nuclear density
at energies around the $a_0(980)$ meson.
 An inspection to the imaginary part of Fig. \ref{fig:R1}
seems to indicate that the peak of its magnitude, corresponding to the 
apparent mass of the $a_0(980)$ resonance, 
 moves to low energies as the
density increases, producing a shift of about $-50$ MeV at $\rho=\rho_0$. This
shift is also visible in the real part by looking at the point where the real
part changes sign. The apparent width measured from the imaginary part of the 
amplitude becomes bigger as the density increases and for $\rho=\rho_0$ 
becomes as large as 200 MeV from an apparent free width of around 90 MeV.
This behaviour is, however, not reproduced in the  
$\pi \eta \to \pi \eta $ amplitude. The first difference one may observe is
the  presence of a larger background, even in free space, which makes it not 
resemble a Breit Wigner resonance so much. For this channel,
the $a_0$ resonance does not move much with the density and the width 
becomes very large already
at small densities, to the point that at densities of the order of one half
$\rho_0$ the resonant shape is practically lost.  Given the fact that the 
$a_0(980)$ meson is usually observed in mass distributions of $\pi \eta$
in the final state of some reaction, the relevant magnitude entering the cross
section of these reactions close to the resonance is the modulus squared of
the $\pi \eta \to \pi \eta $ amplitude \cite{flatte} plotted in 
Fig. \ref{fig:R3}.
 What we observe there is, indeed, that the resonance melts very fast as the 
density increases and at densities of the order of $\rho_0/2$
there is practically no resonant trace left.  The fast disappearance
of this relatively narrow resonance in nuclei is probably one of the most 
striking predictions for this channel.  
\begin{figure}[htb]
 \begin{center}
\includegraphics[height=12.cm,width=12.cm,angle=0] {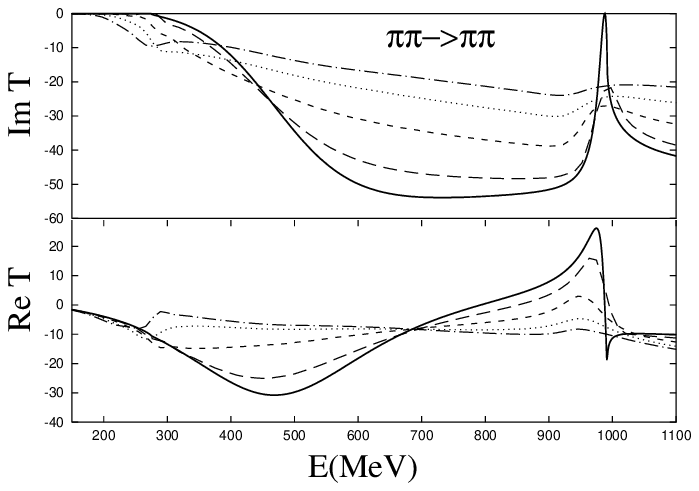}
 \caption{Real and imaginary part of the pion-pion scattering amplitude 
in the I=0 channel as a function of the center of mass energy for different 
nuclear densities. Meaning of the lines as in Fig. 8}
 \label{fig:R4}
 \end{center}
\end{figure}

    The I=0 channel has a richer structure as a function of the density. In
Fig. \ref{fig:R4} we show the amplitude $\pi \pi  \to \pi \pi $ in a range of 
energies from 200 MeV to 1100 MeV. The figure shows results at low energies 
already discussed in \cite{chiang}. As one can see in the figure, there is an
accumulation of strength in the imaginary part below threshold which was first
pointed out in \cite{schuck} and has also been predicted in other approaches 
\cite{aouissat}. The relationship of this increased strength to the enhanced
two pion distribution in $(\pi,2\pi)$ reactions in nuclei \cite{nevio1} at small
invariant masses has been discussed in \cite{rapp2,manolo}, but according to 
the detailed calculation of \cite{manolo} it is not enough to reproduce the
experimental data. 

The intermediate region of energies is quite interesting and no much attention 
has been given to it so far. There one can see a drastic
decrease of the strength of the imaginary part as the density increases. This
reduction could lead to appreciable changes in the two pion production
reactions in nuclei, like the $(\gamma,2 \pi)$ reaction for which 
experiments are already becoming available \cite{krusche}. 
\begin{figure}[htb]
 \begin{center}
\includegraphics[height=12.cm,width=12.cm,angle=0] {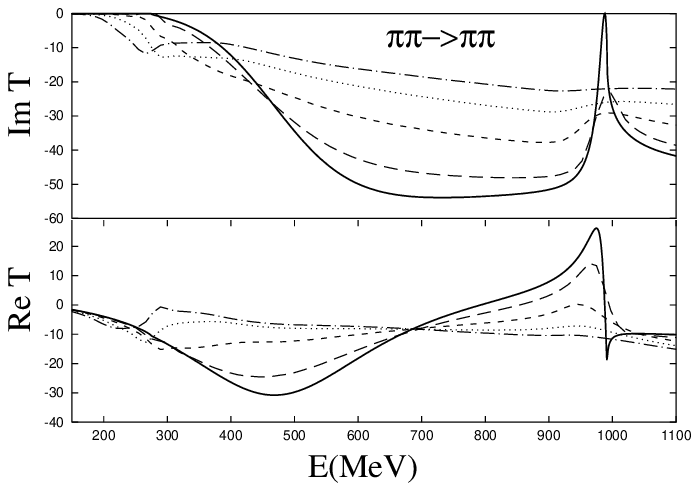}
 \caption{Same as Fig. 11, including $N^*h$ excitations.}
 \label{fig:R5}
 \end{center}
\end{figure}
   The region of the $f_0(980)$ resonance is also interesting. By looking both
at the dip of the imaginary part of the amplitude, as well as to the position
of the zero of the real part, we can see that the position of the resonance
does not change when the density increases.  We observe, however, a gradual
melting of the dip of the imaginary part which comes as an interference
between the background of the $\sigma $ meson contribution and the 
contribution of the $f_0(980)$ resonance. 
\begin{figure}[htb]
 \begin{center}
\includegraphics[height=12.cm,width=12.cm,angle=0] {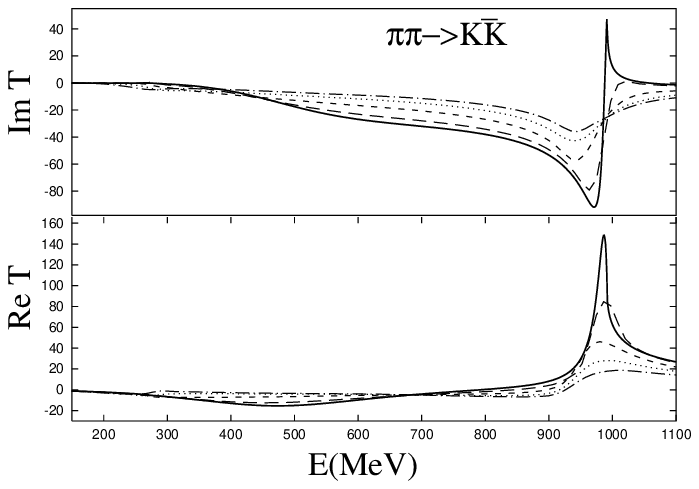}
 \caption{Real and imaginary part of the $\pi\pi\to K\bar{K}$ scattering 
amplitude in the I=0 channel as a function of the center of mass energy 
for different nuclear densities. Meaning of the lines as in Fig. 8}
 \label{fig:R6}
 \end{center}
\end{figure}

It is interesting to discuss what
happens when we introduce the $N^*h$ excitation, which is a novel ingredient
with respect to the approach of \cite{chiang}. In Fig. \ref{fig:R5} we show the results
in which the SU(3) version of the coupling of the $N^*$ to $N \pi \pi$ and 
$N K \bar{K}$ is used.  We can see that the inclusion of this new
ingredient barely modifies the results of the amplitude in all the range of
energies shown. These changes amount to about a 10 per cent increase of the
strength of the imaginary part of the amplitude in the region
around 300 to 400 MeV. We do not show it here, but point out that using the 
SU(2) version of the  $N^*$ coupling to $N \pi \pi$, in which the $N^*$ only 
couples to  $N \pi \pi$, practically does not change the results with respect
 to those in which the $N^*$ is allowed to couple to pions and kaons.

\begin{figure}[htb]
 \begin{center}
\includegraphics[height=12.cm,width=12.cm,angle=0] {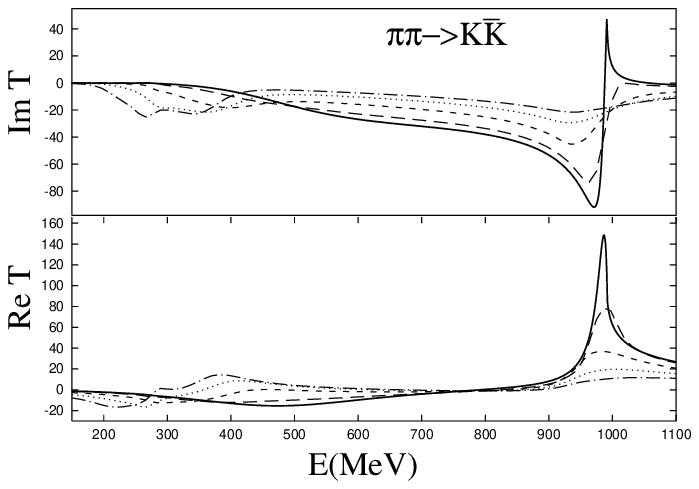}
 \caption{Same as Fig. 13, including $N^*h$ excitations.}
 \label{fig:R7}
 \end{center}
\end{figure}
  The role of the $N^*h$ excitation is more apparent in the 
$K \bar{K} \to \pi \pi$ and $K \bar{K} \to K \bar{K}$ amplitudes.  
In Fig. \ref{fig:R6} we show the 
$K \bar{K} \to \pi \pi$ amplitude for different densities. This amplitude is
better suited than the $\pi \pi \to \pi \pi$ in order to show the $f_0(980)$
resonance because it does not have a large background. 
Notice that in this inelastic channel the $f_0(980)$ shows up
as a Breit Wigner contribution rotated 90 degrees. Thus, the roles of the
real and imaginary parts of the amplitude are interchanged \cite{ollergam}. 
 We can see that as the density increases the position of the resonance barely
moves. The width, however, grows with the density from a free value of 
around 30 MeV to about 100 MeV at  $\rho= \rho_0$. If we  include  the
$N^*h$  contribution in the SU(2) formulation there are no appreciable changes
with respect to those shown in the figure. The results are however quite
different if we include the $N^*h$  contribution in the SU(3) formulation.
These results are shown in Fig. \ref{fig:R7}. We can see there that around 300 
to 400 MeV a resonant like structure develops with the imaginary part showing 
a negative peak and the real part changing fast around a zero value. This is a 
reflection of the $N^*h$ excitation which in this case is magnified because of 
the large coupling of the $N^*h$ excitation to $K \bar{K}$, as we saw in 
section 3. As we discussed there, we found a large coupling, of the order of
$m_K^2/m_\pi^2$ that of the pion, based on a generalization to SU(3) of the
pion coupling. We also saw that there were ambiguities, but any simple 
generalization led to a coupling of this order of magnitude. In spite of this 
huge coupling, we saw no visible effects in the $\pi\pi\to\pi\pi$ amplitude.  
Here it shows clearly in a large medium change of the $\pi \pi \to K \bar{K}$ 
amplitude. The effects in the $K\bar{K} \to K\bar{K}$ amplitude are even more 
pronounced. 
\begin{figure}[htb]
 \begin{center}
\includegraphics[height=12.cm,width=12.cm,angle=0] {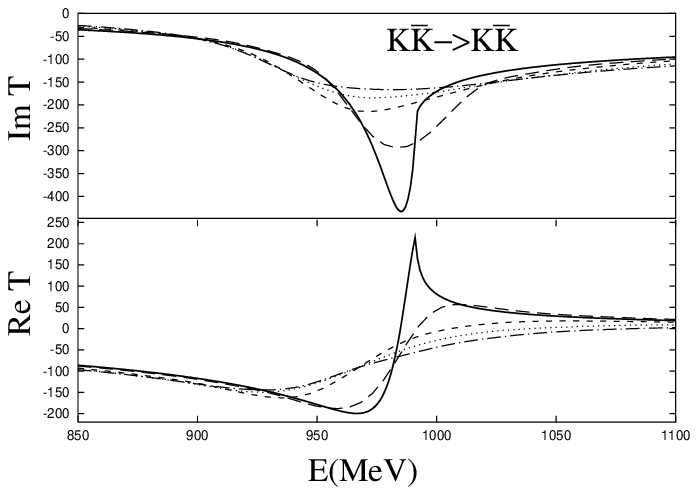}
 \caption{Real and imaginary part of the kaon-kaon scattering amplitude 
in the I=0 channel as a function of the center of mass energy for different 
nuclear densities. Meaning of the lines as in Fig. 8}
 \label{fig:R8}
 \end{center}
\end{figure}

In Fig. \ref{fig:R8} we show the $K \bar{K} \to  K \bar{K}$ in the energy 
region around the $f_0$ resonance. The density effects around the pole can be 
appreciated better and one can see that even at $\rho= \rho_0$ the shape of the 
resonance is not lost, but the width  increases to about 100 MeV at $\rho_0$.

  As we have said, there are some elements of uncertainty tied to the
extrapolation of the $N^*$ coupling to kaons, and it would be worth
trying to find some observable consequences of the assumptions made. Since the
larger effects are seen in a region where the kaons are far off shell, one can 
only hope to observe indirect effects on reactions were the kaons appear as
intermediate states. The large values obtained in the amplitudes will certainly
be softened  by the small weight of a $K\bar{K}$ propagator where the two kaons
are quite off shell, so one should not expect drastic changes. Yet, even 
moderate changes might be relevant in some processes like the $(\pi, 2\pi)$ 
reaction in nuclei, where it was shown in \cite{manolo} that there were large 
cancellations between terms to give a final result smaller than  the 
contribution of individual terms, such that any small changes in one of them 
might alter the final balance. The finding of indirect evidence of this directly
unobservable $N^*$ coupling to N and kaons would be an important test of 
particle symmetries.  

   In any case the interesting medium effects found here, independent of the
still unknown couplings, would certainly call for devoted experiments from
which we could learn more about the nature of the scalar resonances and the
way the meson meson interaction is changed in a medium. Reactions like
$\gamma p \to \pi \pi p$ have already been suggested as a means to
observe the scalar resonances \cite{eugenio}. Their extension using nuclear
targets is certainly feasible and, together with other experiments, should be
encouraged.

\section{Conclusions} 
In  section 3 we addressed the question of new contributions to
the $\pi \pi $ scattering in a nuclear medium beyond those already considered
in other approaches. One of the terms considered in which a nucleon loop is
attached to the four meson vertex was found to be zero for symmetric nuclear
matter. Other possible mechanisms which would simulate  a three sigma
vertex coupling were also estimated to be much weaker than previously 
suggested. These 
results would further strengthen those obtained in \cite{chiang,manolo},
which would in turn mean that the experimental problem of the enhanced
invariant $\pi \pi$ mass close to threshold would not  be solved yet.

The main topic of the present paper has been the discussion of the
renormalization of the properties of the scalar meson resonances, concretely
the $f_0 (980)$ and the $a_0 (980)$ resonances, in the nuclear medium. The 
renormalization
required the use of the kaon selfenergy for which we have used a recent one
deduced from chiral Lagrangians and which is consistent with the information
of kaonic atoms. We have systematically tried to use the chiral unitary 
formalism in the different aspects of the problem, be the generation of the 
resonances through the meson meson interaction given by
the chiral meson Lagrangians, or the meson baryon interaction, which for the
most delicate case, the one of the $K^-$, is also obtained by means of a
nonperturbative chiral approach.
The results obtained are interesting, we do not observe an appreciable
change of the position of either resonance. However, the widths are
substantially changed. In the case of the $f_0(980)$ resonance the width
passes from    30 MeV in the free case to about 100 MeV  at normal nuclear 
matter.  In the case of the $a_0(980) $ resonance the width grows so fast with
density that even at $\rho_0/2$ there is practically no trace of the resonance.
The next step should be the search for these effects in nuclear experiments
which  can help shed new light  on the nature of these resonances and 
the behaviour of kaons in nuclear matter.

\vspace{3cm}
\noindent
Acknowledgments:

We would like to thank A. Ramos for providing us with her codes to 
calculate kaonic spectral functions. Useful discussions with N. Grion,
T. Hatsuda, G. Chanfray, R. Rapp, and J. Wambach are also acknowledged. 
This work is partly supported by DGICYT contract no. PB 96-0753 and by 
the EEC-TMR Program, EURODAPHNE, Contract No. ERBFMRX-CT98-0169.

\end{document}